\documentclass[12pt]{article}
\usepackage[a4paper, margin=1in]{geometry} 
\usepackage{setspace}
\usepackage[version=3]{mhchem} 
\usepackage{amsmath}
\usepackage{ulem}
\usepackage{authblk}
\usepackage{graphicx}
\usepackage{cite}
\usepackage[table]{xcolor}
\usepackage[utf8]{inputenc}
\usepackage{etoolbox}
\usepackage[T1]{fontenc}
\usepackage[colorlinks, linkcolor=blue,citecolor=blue, urlcolor=blue,bookmarks=true]{hyperref} 

\hypersetup{
colorlinks = true, 
urlcolor   = black,
linkcolor  = black, 
citecolor  = black 
}

\usepackage{xr}
\usepackage{xcite}
\makeatletter
\newcommand*{\addFileDependency}[1]{
\typeout{(#1)}
\@addtofilelist{#1}
\IfFileExists{#1}{}{\typeout{No file #1.}}
}\makeatother

\definecolor{darkgreen}{rgb}{0.0, 0.42, 0.24}

\newcommand*\revision[1]{\color{black}}

\definecolor{hcyan}{rgb}{0.0, 0.8, 0.8}
\definecolor{dorange}{rgb}{0.8, 0.4, 0.0}

\newcommand{\hh}{H$_2$~}

\title{Towards viable H$_2$ storage in Ca decorated low-dimensional materials with insights from reference quantum Monte Carlo
}

\author[1,2]{Yasmine S. Al-Hamdani}
\author[1,2]{Dario Alfè}
\author[1,2]{Andrea Zen*}

\affil[1]{Department of Earth Sciences, University College London, London WC1E 6BT, United Kingdom}
\affil[2]{Dipartimento di Fisica Ettore Pancini, Università di Napoli Federico II, Monte S. Angelo, I-80126 Napoli, Italy}

\begin{document}
\footnotetext[1]{Corresponding author email: andrea.zen@unina.it}

\maketitle

\begin{abstract}
Hydrogen technology is set to be a key energy alternative for mitigating pollution and reducing CO$_2$ emissions. 
However, the current storage mechanism of hydrogen molecules in carbon fibre tanks detracts from the fuel economy of hydrogen in mobile applications, necessitating the development of alternative storage mechanisms. Adsorbing hydrogen in its molecular form (H$_2$) at typical operating conditions of proton exchange membranes can potentially meet storage requirements. However, H$_2$ is the smallest molecule with only two electrons and therefore it has very limited propensity to physisorb in a material within the binding energy window of $-0.2$ to $-0.4$ eV that is suitable for storage. Calcium atom decorators on graphene have previously shown promise for tunable H$_2$ binding, but the system is thermodynamically unstable toward the formation of calcium hydride. Moreover, the absolute adsorption of H$_2$ is challenging to predict accurately and is typically overestimated with van der Waals inclusive density functional approximations. In this work, we perform \textit{state-of-the-art} fixed-node diffusion Monte Carlo alongside a selection of density functional approximations for two strategies of anchoring Ca: (i) Ca on boron doped graphene and (ii) Ca inside carbon nanotubes. We predict reliable Ca and H$_2$ binding energies, and establish that Ca is anchored inside carbon nanotubes and on boron doped graphene, while boosting the H$_2$ adsorption energy. Importantly, the H$_2$ adsorption energy is found to be improved by the anchoring strategies, with the energy inside a Ca decorated carbon nanotube reaching the viable storage window.  The reference DMC binding energies provide much-needed benchmarks for developing data-driven methods and guiding experiment in the systematic design of hydrogen storage materials. 
\end{abstract}

\section*{Introduction}\label{sec:introduction}
\hh technology is one of the key strategies for mitigating greenhouse gas emissions as \hh molecules are converted into energy and water using proton exchange membranes (PEMs), delivering a high energy density with no local CO$_2$ emissions.  However, novel storage materials are needed to propel \hh technology, particularly for mobile applications. Currently, expensive carbon fibre tanks are used to contain pressurized \hh gas up to $\sim700$ bar \cite{Usman2022HydrogenStatus,RahmatPoudineh2023HydrogenFuture}. Pressurizing \hh to this extent is an energy intensive process and detracts significantly from the energy efficiency of using \hh. Alternative lower pressure storage solutions are sought to improve the energy efficiency of using \hh and minimize the cost of storage. To this end, the adsorption energy of \hh molecules plays a major role in the viability of \hh storage materials. Strong dissociative adsorption of \hh as H atoms requires considerable desorption or recombination energies to release \hh for PEMs. Instead, adsorbing molecular \hh in an energy window of approximately $-0.2$ to $-0.4$ eV  per \hh molecule in low-dimensional materials, has been theorized to correspond to the operating conditions of a typical PEM \cite{alhamdani2023prm,Bhatia2006OptimumStorage,Li2003TheoreticalNanostructures}. 

The challenge for meeting the target adsorption energy window for \hh lies in its small, two-electron, and non-polar nature. Many works, especially using density functional theory (DFT), have focused on this challenge with some applying a high-throughput approach to tackle the many thousands of low-dimensional materials being synthesized or hypothesized \cite{Bobbitt2016High-ThroughputTemperature,Bobbitt2019MolecularStorage,Colon2014High-ThroughputTemperature,Gao2024ExperimentallyStorage}, and others applying a bottom-up approach focusing on the mechanistic details \cite{Ataca2009,Srinivas2009HighCaC6,Liu2011,Zhou2012,Seenithurai2014Li-decoratedStudy,Wong2014,Qiu2014AStorage,Wen2017,al-hamdani2017cnt,alhamdani2023prm,alhamdani2023jcp}.  In our previous work we applied the latter approach and also reported beyond-DFT adsorption energies for a selection of group 1 and group 2 metal decorated graphene substrates \cite{alhamdani2023jcp,alhamdani2023prm,al-hamdani2017cnt}. It has been shown that group 1 metal adatoms boost the \hh adsorption energy well beyond the weak adsorption on pristine graphene, through static polarization \cite{alhamdani2023prm}. This form of interaction is weakened with each additional \hh molecule and group 1 metal decorated graphene is therefore at the edge of being a useful storage material. On the other hand, Ca and Sr decorated graphene, has been shown to facilitate the adsorption of multiple \hh molecules through a weak covalent interaction known as Kubas bonding\cite{alhamdani2023prm,Kubas2001,Kubas2007DihydrogenMolecules}. This form of bonding is only exhibited where an occupied metal d-state is available to overlap with the anti-bonding \hh 1$\sigma*$ state.  However, due to the intricate balance of physical interactions including long-range dispersion forces, static polarization, charge transfer, and weak covalent bonding, the \hh adsorption energy is very sensitive to the computational method that is applied, and in particular to the delocalization error in density functional approximations (DFAs) \cite{Bajdich2010}. Meanwhile, state-of-the-art diffusion Monte Carlo (DMC) is a wavefunction based method that is free of the delocalization error and accounts for full many-body dispersion interactions.  DMC adsorption energies are therefore accurate references that can be used as a guide in DFT modelling to arrive at reliable insights. For example, we have shown using DMC that Kubas binding \hh is feasible on Ca decorated graphene \cite{alhamdani2023jcp}. 

An important current challenge is to ensure the stability of the single atom Ca decorator.  It has previously been shown that Ca has a low barrier to diffuse across the graphene surface and that the system is unstable towards the formation of calcium hydride and graphite \cite{Wood2011Ca-intercalatedGraphite,alhamdani2023prm}. Several strategies have previously been considered for anchoring Ca adatoms and two     mechanisms that we consider here are: (i) to increase the Ca-substrate interaction through boron doping of graphene, and (ii) to increase the physical barrier for Ca to agglomerate by adsorbing inside carbon nanotubes (CNTs).  First, boron doping in graphene reduces the electron density relative to pristine graphene and therefore strengthens the electron transfer from Ca to the surface. Second, CNTs typically adsorb atoms and molecules more strongly than pristine graphene \cite{al-hamdani2017cnt} and thanks to their lower dimensionality, CNTs have a more closed structure that is expected to help segregate adatoms such as Ca.  Given the difficulty of accurately capturing the \hh interaction, we apply a selection of widely-used and well-performing DFAs alongside DMC, to predict the anchoring of Ca adatoms and the impact of these strategies on \hh adsorption. 
In the following, we detail the system setup and computational methods in Section \ref{sec:methods}, we report the results for Ca decorated BGr in Section \ref{sec:bgr} and for Ca decorated CNTs in Section \ref{sec:cnt}. We provide a brief discussion in Section \ref{discussion} and conclude in Section \ref{conc}.

\section{Methods}\label{sec:methods}
\subsection{System setup}
The boron doped graphene (BGr) system is made by the substitution of a single carbon atom in a ($5\times5$) unit cell of graphene, with a 20.0 Å vacuum in the direction perpendicular to the sheet as shown in Fig.~\ref{fig:2d}. CNT based systems have a vacuum of 10.0 Å in perpendicular directions to the CNT and approximate length of $8.5$ Å. Structures can be found in the Supplemental Materials (SM)~\cite{SM}. The binding energy ($E_b^{H_2}$) of a hydrogen molecule is computed as:
\begin{equation}
E_b^{H_2} = ( E_{nH_2+substrate} - E_{substrate} - nE_{H_2} ) / n 
\end{equation}
where $E_{nH_2+substrate}$ is the total energy of $n$ \hh molecules adsorbed on a substrate, $E_{substrate}$ is the total energy of a substrate, and $E_{H_2}$ is the total energy of an isolated \hh molecule in the gas phase. The substrate system includes Ca, where present, for computing $E_b^{H_2}$. The binding energy of a Ca adatom, $E_b^{Ca}$ is computed as:
\begin{equation}
E_b^{Ca} = E_{Ca+substrate} - E_{substrate} - E_{Ca}  ,
\end{equation}
where $E_{Ca}$ is total energy of an isolated Ca atom in the gas phase. Each system is fully relaxed using PBE+D3\cite{Perdew1996,Grimme2010a} and a consistent unit cell is used across all terms for each binding energy. 

\subsection{Structure search}
The potential energy surface is typically shallow for physisorbed H$_2$ molecules and therefore a random structure search was used to generate 50 configurations of \hh in the CNT(10,0) substrate and optimized with PBE+D3 \cite{Perdew1996,Grimme2010a} and the CP2K code \cite{Kuhne2020}. The initial \hh molecules were generated to be within 3.5 Å of the Ca atom and on the inside of the CNT, with random orientation. The absolute energy difference among the 20 lowest energy structures was less than $\sim15$ meV. Ten lowest energy structures were optimized using the numerical settings in VASP v.6.2.1 \cite{Kresse1993,Kresse1996,Kresse1996a} given in Section~\ref{dftsetup}. The lowest energy Ca-H$_2$ configuration in CNT(10,0) was used as the starting geometry for CNT(8,0), CNT(5,5) and CNT(6,6).   

\begin{figure}[htbp]
    \centering
    \includegraphics[width=0.5\linewidth]{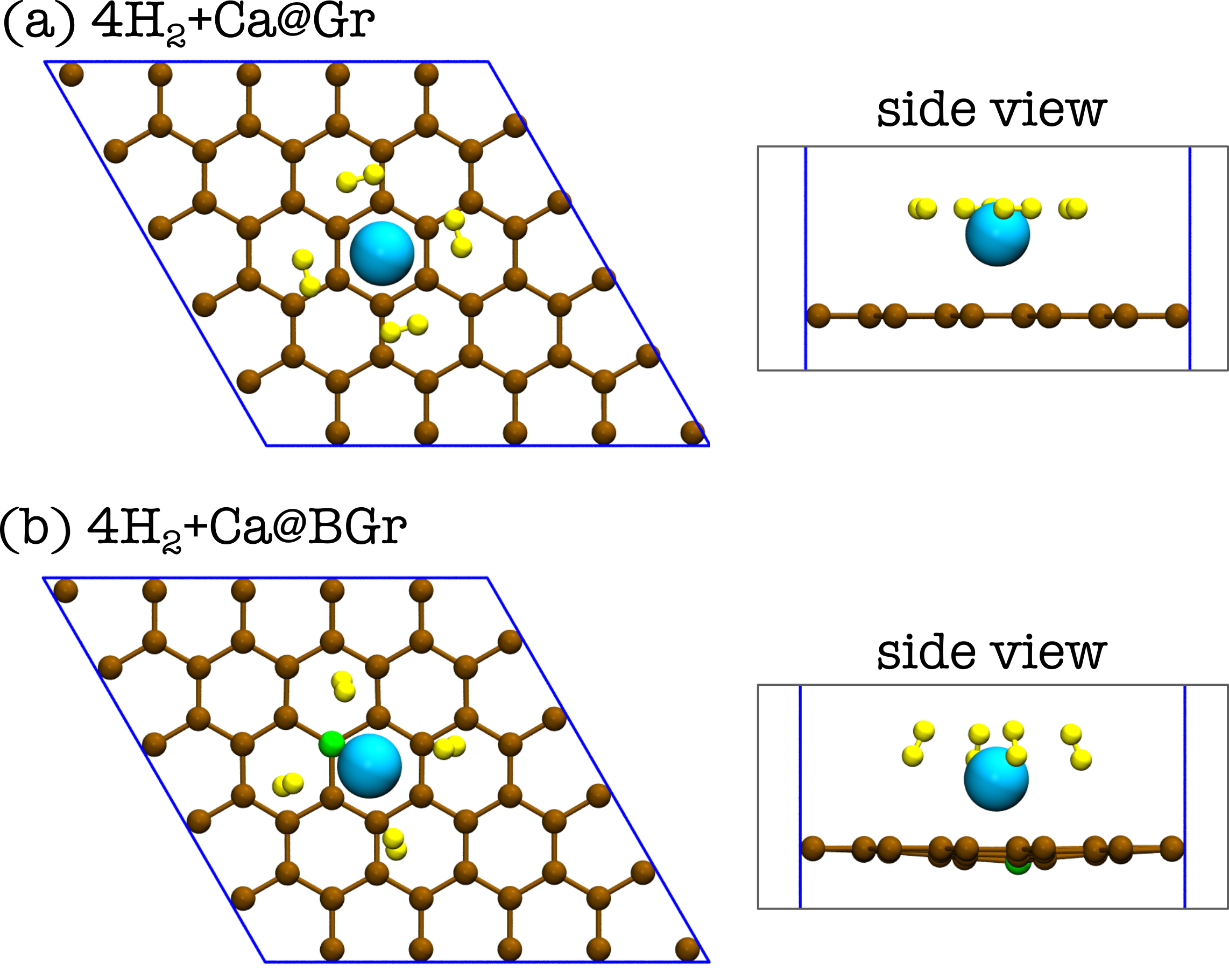}
    \caption{Configurations for 4H$_2$ adsorption on 2D materials with H atoms in yellow, Ca in blue, C in brown and B in green: (a) Ca decorated graphene (Ca@Gr) and (b) Ca decorated B-doped graphene (Ca@BGr). The configurations are obtained from PBE+D3 geometry optimizations. }\label{fig:2d}
\end{figure}

We geometry optimized Ca at symmetrically distinct rings on the surface of BGr to find the lowest energy adsorption site, as well starting with atom-top initial configurations. The lowest energy configuration has Ca on the BC$_5$ ring with Ca shifted towards B. Similarly, we geometry optimized upright, flat, and radial orientations of one and four H$_2$ molecules around the Ca atom on BGr (on the lowest energy Ca@BGr structure), to find the lowest energy configuration. The adsorption of 4H$_2$ molecules does not change the site of Ca adsorption on BGr and the final 4H$_2$+Ca@BGr geometry can be seen in Fig.~\ref{fig:2d}. 

\subsection{DFT setup}\label{dftsetup}
Standard PAW potentials \cite{Kresse1994,Kresse1999} are used in the VASP v.6.2.1 code, with a 500 eV plane-wave cut-off and the configurations have been optimized with PBE+D3 (zero damping and no three-body ATM terms) to forces smaller than 0.005 eV Å$^{-1}$ and an energy criterion of $10^{-6}$ eV. A $9\times9\times1$ and a $1\times1\times7$ $\Gamma$ centered \textbf{k}-point grid is used in DFT calculations for the 2d and 1d systems, respectively. Gas phase H$_2$ and Ca energies are computed in the corresponding unit cells for each binding energy calculation.  $E_b^{H_2}$ is converged to within 1 meV at the chosen \textbf{k}-point grids with respect to fully converged grids. The lowest energy structure found with PBE+D3 for each system is fixed for subsequent benchmarking with a selection of DFAs (optB86b-vdW \cite{Klime2011}, r2SCAN \cite{Furness2020}, r2SCAN+rVV10 \cite{Furness2020,Sabatini2013}, PBE+MBD \cite{Perdew1996,Tkatchenko2012,Ambrosetti2014}, PBE+MBD-FI \cite{Perdew1996,Gould2016,Gould2016b}, rev-vdW-DF2 \cite{Hamada2014}, PBE, and the LDA) and DMC. 

\subsection{QMC setup}
We used Quantum Espresso v6.8 \cite{Giannozzi2009,Giannozzi2017} to compute the LDA orbitals as input for single Slater determinant QMC computations in the QMCPACK (v.3.9) code \cite{Kim2018}. Non-spin polarized LDA orbitals were computed using 400 Ry plane-wave cutoffs in combination with ccECP potentials for H, Ca, C and B atoms \cite{Bennett2018}. The planewave orbitals were localized using B-splines\cite{Alfe2004} and a mesh factor of 0.5 in QMCPACK. The workflows were managed by the NEXUS python package \cite{Krogel2016}. For a more efficient Brillouin sampling, the Baldereschi point was used in reciprocal space for Gr and BGr based systems \cite{Baldereschi1973}. Full details of the QMC twists and supercells used can be found in the SM~\cite{SM}. The spin polarization contribution was estimated at the LDA level for BGr based systems. There is no spin-polarization contribution to the CNT(6,6) based systems. One and two-body Jastrow factors have been optimized for all systems using the OneShiftOnly algorithm as implemented in QMCPACK. The determinant localization approximation \cite{Zen2019,DellaPia2025} was used for all DMC calculations along with a careful time-step convergence (see SM for details~\cite{SM}). The QMC calculations are prone to finite-effects that stem from explicit electrons and this has been estimated at the DFT level with the KZK correction \cite{Kwee2008}. Numerical details can be found in the SM~\cite{SM} while we report DMC values that carry the smallest KZK corrections in Section~\ref{sec:results}.

\section{Results}\label{sec:results}

\subsection{Adsorbing multiple H$_2$ molecules on Ca decorated boron doped graphene}\label{sec:bgr}

We have previously found that the adsorption of \hh on pristine graphene is very weak, at $-20\pm3$ meV from DMC \cite{alhamdani2023jcp}. We also found that decorating pristine graphene with Ca, boosts the \hh binding energy by \textit{ca.} 90 meV \cite{alhamdani2023jcp}. While this is still outside of the viable binding energy window for \hh storage, the bonding mechanism involves a Ca $3d$ to \hh $1\sigma^*$ (Kubas) interaction which makes it energetically favorable to bind four \hh molecules due to the symmetry of the states. This is different to static polarizable adsorption of \hh as exhibited by group 1 metal decorated graphene, \textit{e.g.} Li@Gr and Na@Gr, where the maximal binding energy occurs for the first \hh molecule and it decreases for each additional \hh \cite{alhamdani2023prm}.  
However, Ca has a low barrier to diffuse across pristine graphene and previous works indicate that it is thermodynamically unstable towards the formation of calcium hydride and graphite in the presence of \hh \cite{Wood2011Ca-intercalatedGraphite}. First, we report the binding energy of Ca on boron doped graphene (Ca@BGr) to establish whether boron doping makes Ca adsorb on graphene more strongly. Second, we predict the adsorption energy of 4H$_2$+Ca@BGr with a selection of widely-used DFAs and DMC to understand the impact on the \hh interaction.
\begin{figure}[htbp]
    \centering
    \includegraphics[width=0.6\linewidth]{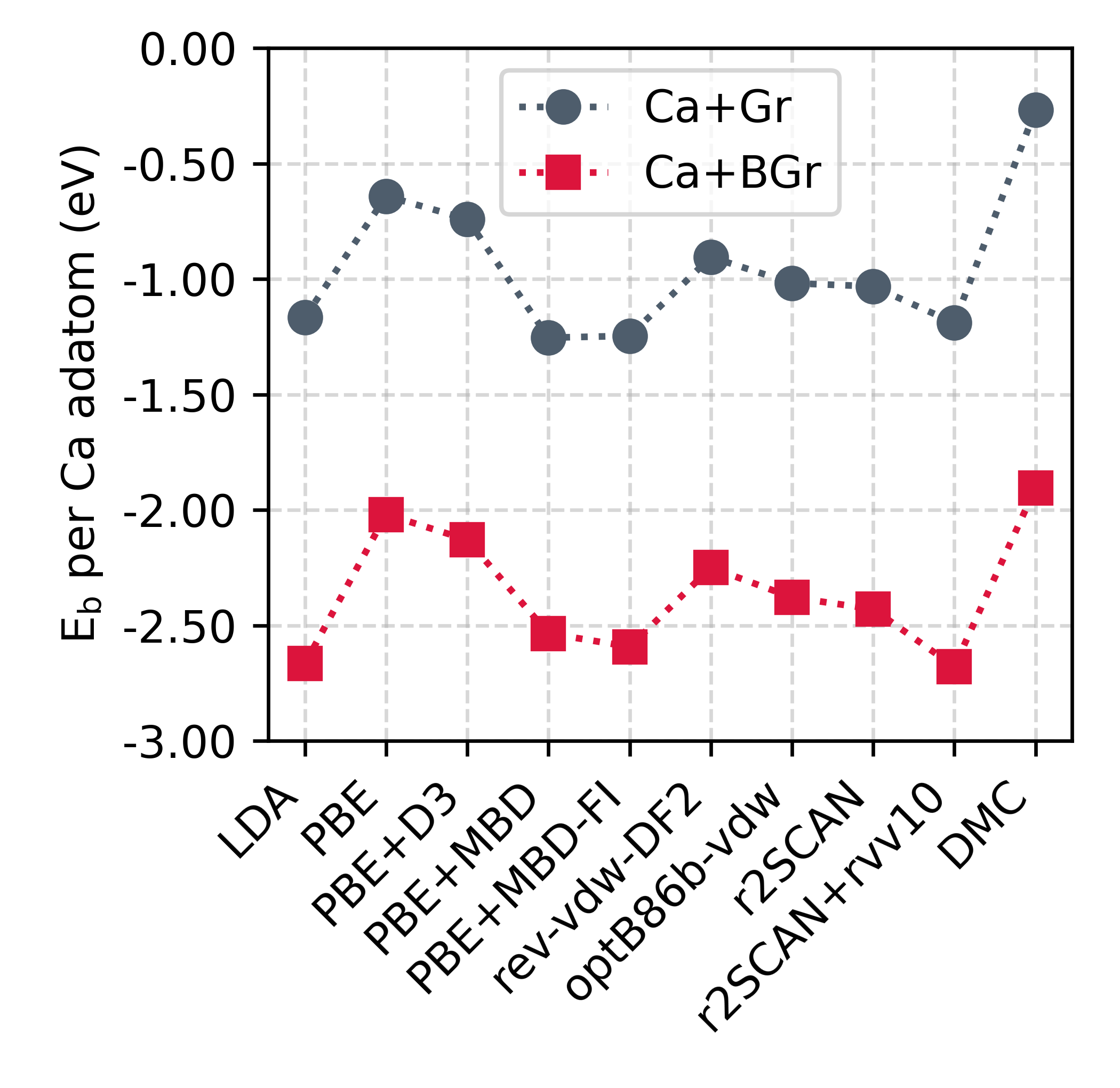}\label{fig:cagr}
    \caption{Binding energy of Ca on graphene in gray and B-doped graphene in red, across a selection of DFAs and DMC.}
 \end{figure}   
\begin{figure}[htbp]
    \centering
    \includegraphics[width=0.6\linewidth]    {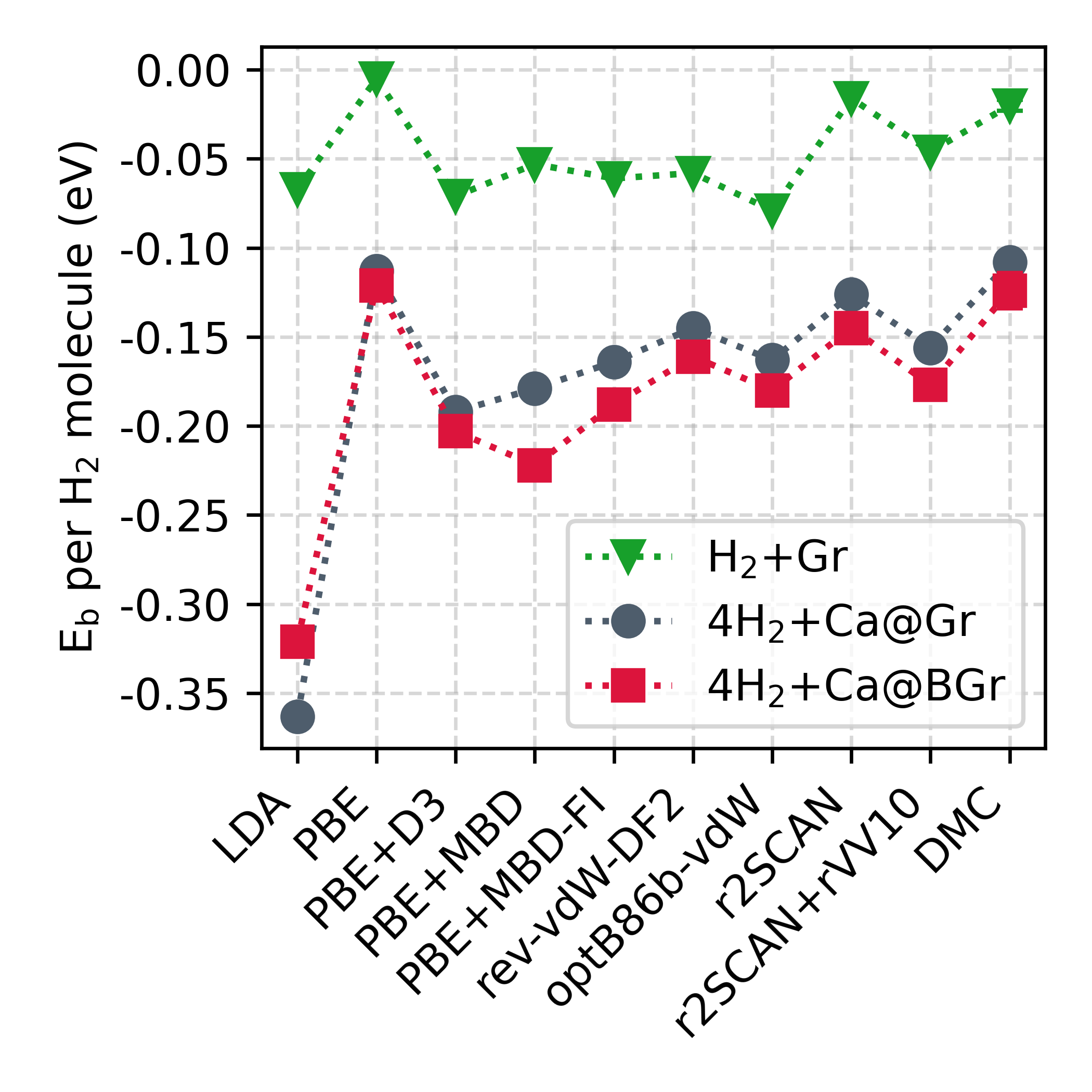}
    \caption{Binding energy of 4H$_2$ per H$_2$ molecule on Ca@Gr in gray and Ca@BGr in red, across a selection of DFAs and DMC. The binding energy of a single H$_2$ molecule on pristine graphene is also shown with green triangles.}\label{fig:h2cagr}
\end{figure}

We can see from Fig.\ref{fig:cagr} that B doping graphene substantially strengthens the adsorption of Ca on the substrate. An increase of $\sim1.5$ eV in the Ca adsorption strength is predicted consistently by all of the electronic structure methods we apply, including DMC. Furthermore, we computed the PBE+D3 energy barrier for the diffusion of Ca to the nearest carbon-only ring in the surface, using the nudged-elastic-band (NEB) method (full details are given in the SM~\cite{SM}). On pristine graphene, the Ca diffusion barrier was previously found to be 0.13 eV using the same methodology \cite{alhamdani2023prm}. We find here that there is a monotonic energy increase of $0.28$ eV for the Ca to reach the carbon-only ring in BGr, indicating an improvement in the localization of the Ca adatom. The indication on the whole is that increased stability is brought by the markedly stronger binding energy of Ca on BGr. Our findings support the use of BGr for anchoring Ca atoms and a similar effect can be expected for other metal adatoms where the metal adatom binding is charge transfer mediated. It is noteworthy that for Ca@Gr, the DFAs we consider overestimate the adsorption of Ca by $\sim0.5$ eV. Meanwhile there is much better agreement for Ca@BGr with PBE and PBE+D3 (within $\sim0.3$ eV of the DMC reference). It can also be seen that PBE and r2SCAN predictions are made worse by the inclusion of dispersion interactions, indicating that these systems are prone to large delocalization errors at the GGA and meta-GGA level. 

Having established that Ca is strongly adsorbed on BGr, we consider the adsorption of four \hh molecules on Ca@BGr. The adsorption motif was optimized with the PBE+D3 functional and fixed for all other electronic structure methods that we apply. This is necessary in order to compare the interaction strengths with DMC, for which it is not feasible to optimize the geometry. For context, we also report the \hh adsorption energy on pristine graphene (H$_2$+Gr) and 4H$_2$+Ca@Gr in Fig.~\ref{fig:h2cagr} from previous work \cite{alhamdani2023jcp}. We compute these systems here with the same selection of DFAs and settings as given in Section \ref{sec:methods} for consistency. It is evident that pristine graphene can only very weakly physisorb a \hh molecule and it would not be a viable storage material on its own, while introducing a Ca adatom boosts the \hh interaction by $\sim100$ meV. Moreover, it has previously been established that \hh is bound via Kubas bonding on Ca decorated graphene, such that four \hh molecules are adsorbed around a Ca atom. While anchoring Ca with B doping, it can be seen from Fig.~\ref{fig:h2cagr} that DMC and all DFAs considered, with the exception of LDA, predict a weak boost of 10-20 meV in adsorption strength per \hh molecule. Therefore, B doping graphene can be used to stabilize Ca adatoms without deteriorating the \hh adsorption. The consistency among GGA and meta-GGA functionals for the order of adsorption on Ca@Gr and Ca@BGr surfaces is a positive indication for the use of these methods, although the predicted \hh adsorption energy varies by up to $\sim70$ meV even among DFAs that account for some dispersion interactions. Indeed, PBE+D3 and PBE+MBD predictions fall marginally within the viable window of binding energy for \hh storage but without a reference it would not be possible to ascertain which prediction is best. Thanks to the DMC reference it can be seen that PBE, r2SCAN and rev-vdW-DF2 predict the most accurate adsorption energies. Given the lack of long-range dispersion interactions in PBE and r2SCAN, we suggest that a more physically motivated choice is the rev-vdW-DF2 functional. 

\subsection{Viable H$_2$ binding strength in Ca decorated carbon nanotubes}\label{sec:cnt}

\begin{figure}[htbp]
    \centering
    \includegraphics[width=0.5\linewidth]{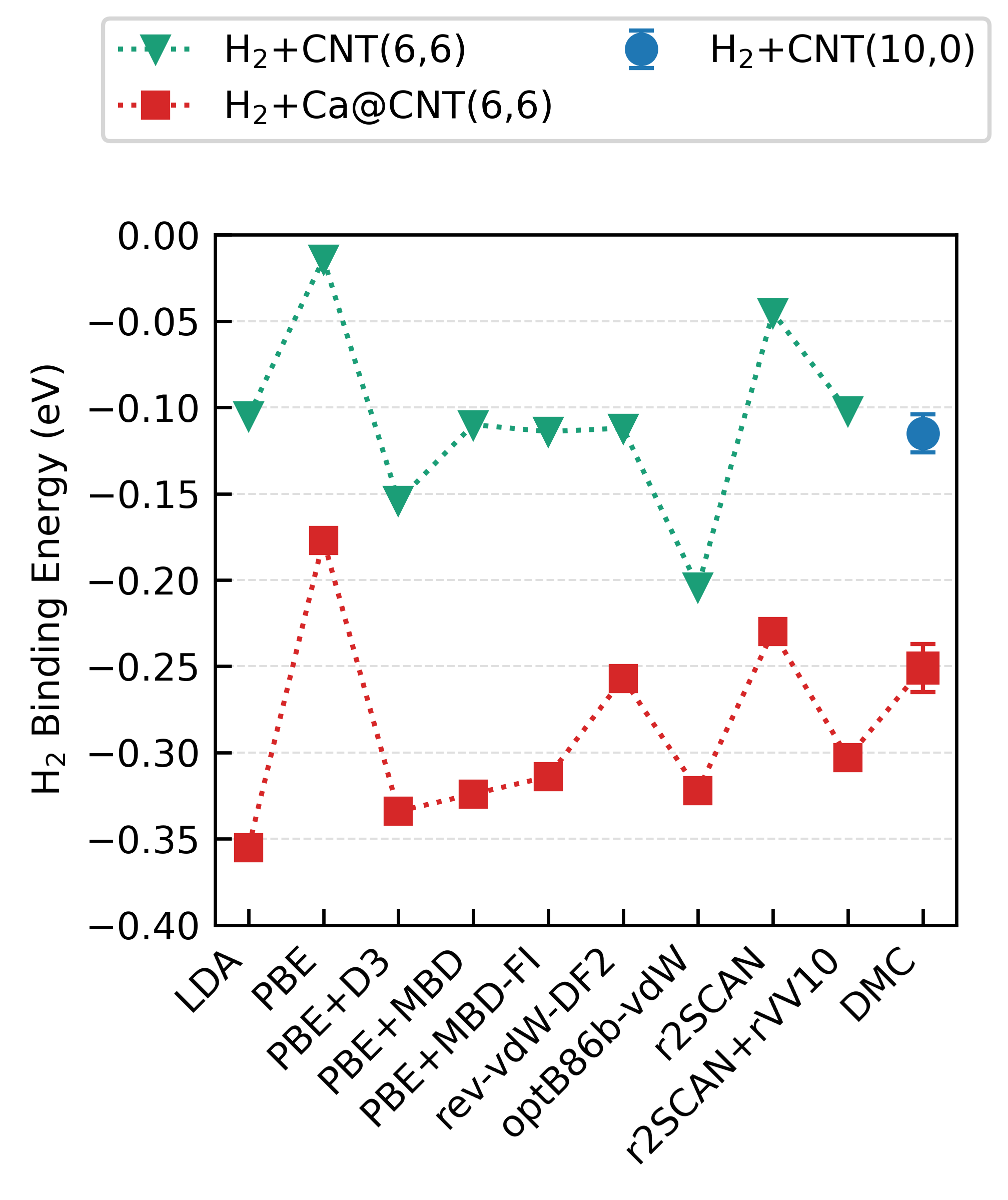}
    \caption{The H$_2$ binding energy with a selection of DFAs inside pristine CNT(6,6) in green triangles, and inside Ca decorated CNT(6,6) in red squares. The DMC binding energy of H$_2$ inside pristine CNT(10,0) from Ref.~\citenum{al-hamdani2017cnt} is shown with a blue circle for reference.}\label{fig:h2cnt}
\end{figure}

\begin{figure}[htbp]
    \centering
    \includegraphics[width=0.5\linewidth]{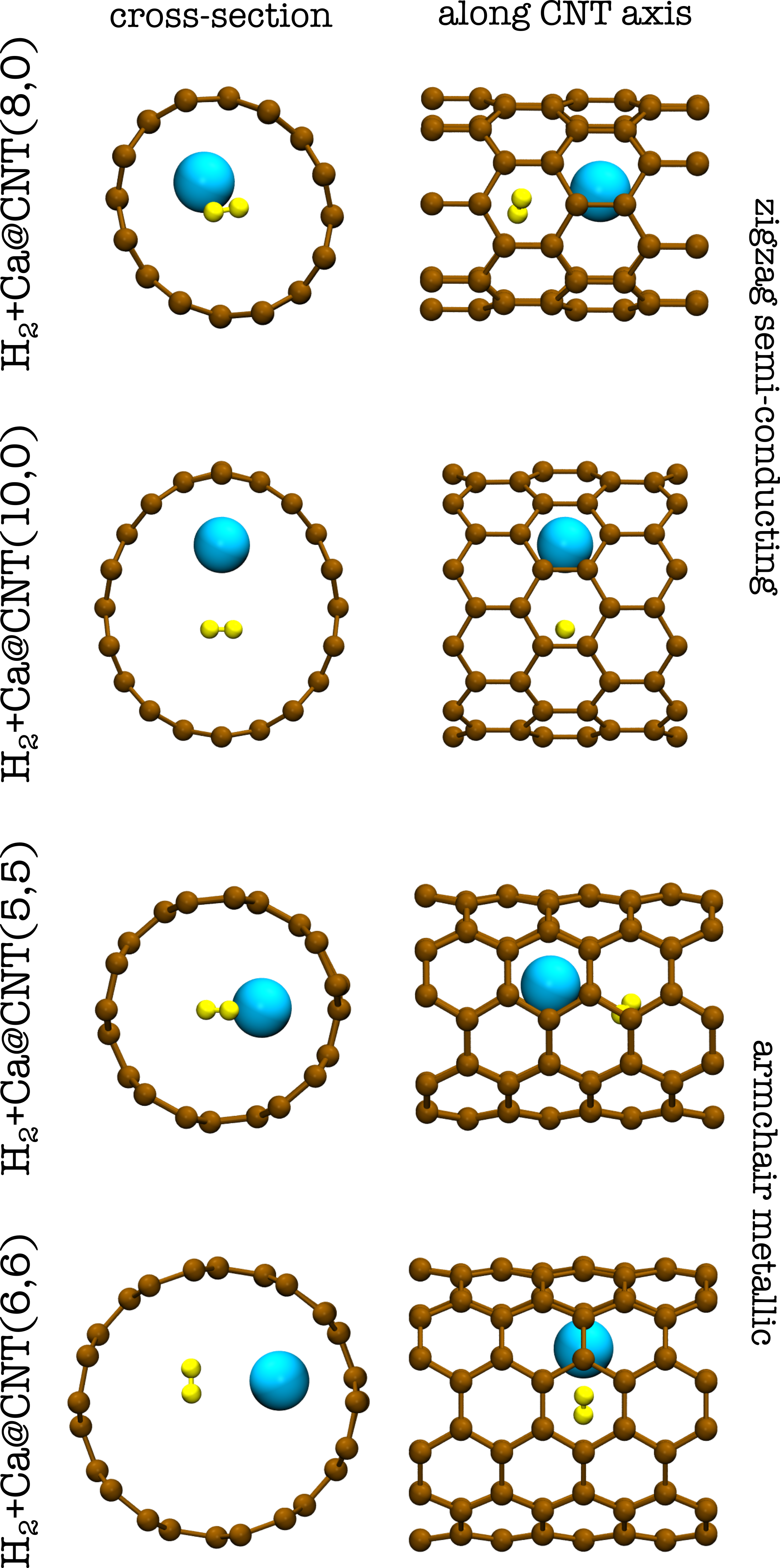}
    \caption{H$_2$+Ca@CNT configurations obtained from RSS and PBE+D3 geometry optimizations. From the top: zigzag semi-conducting CNTs (8,0) and (10,0) with diameters 6.26 and 7.83 Å, respectively, and armchair metallic CNTs (5,5) and (6,6) with 6.74 and 8.14 Å diameters, respectively. All configurations are treated in full periodic boundary conditions (unit cell boundaries not shown) with a vacuum of 10.0 Å perpendicular to the CNT tube axis and the nearest Ca-Ca distance along the CNT axis is 8.5 Å. H atoms in yellow, C in brown, and Ca in blue.}\label{fig:1d}
\end{figure}

Carbon nanotubes (CNTs) are readily synthesized materials and two key characteristics are the tube diameter and the semi-conducting (zigzag) or metallic (armchair) nature of the tube. There has been a longstanding interest in exploiting the open structure of CNTs for gas storage but the controlled synthesis of specific CNTs remains an experimentally challenging process. It is therefore useful to gain insight from computational modeling on the effect of the CNT diameter and electronic state on the adsorption energy of \hh. Many previous works have focused on this for clean and unmodified CNTs showing that \hh generally binds most effectively in CNTs with a diameter of 7 to 10 Å, irrespective of whether a metallic or semi-conducting CNT is used \cite{Cheng2005MolecularNanotubes,Yin2000MolecularArrays,Kagita2012QuantumNanotubes,Lee2012InfluenceBehaviors,Lyu2020AnAdsorption,Chen2004RapidNanotubes}.
We have previously shown using DMC that the interaction energy of \hh inside a semi-conducting CNT(10,0) with 8.1 Å diameter is $-115\pm11$ meV \cite{al-hamdani2017cnt}. 
This is considerably stronger than binding on pristine graphene but still weaker than the ideal binding energy window for \hh storage. Moreover, many widely-used DFAs were shown to overestimate \hh adsorption inside the CNT, due to the accumulation of errors in medium-range electron correlation. This source of inaccuracy can be expected for any molecule under confinement and therefore, makes it difficult to rely on a DFA without a reliable reference of the adsorption energy. Here, we computed the DMC adsorption energy for a \hh molecule inside a Ca decorated armchair CNT(6,6) as shown in Fig.~\ref{fig:1d}, alongside a selection of widely-used DFAs. Note that CNT(6,6) has a diameter only $\sim0.3$ Å larger than CNT(10,0), but the system is non-spin polarized upon the adsorption of Ca, unlike CNT(10,0). As such, it is more memory-efficient to compute the wavefunction for \hh molecule in Ca@CNT(6,6) with DMC. Note that the LDA, PBE, PBE+D3 and PBE+MBD \hh binding energies in pristine CNT(10,0) reported previously\cite{al-hamdani2017cnt} are within 10 meV of the corresponding binding energies in CNT(6,6). Assuming a similar consistency at the DMC level, the DMC \hh binding energy for pristine CNT(10,0) and Ca@CNT(6,6), as shown in Fig.~\ref{fig:h2cnt}, give an approximate indication about the effect of Ca decoration from DMC. It can be seen from Fig.~\ref{fig:h2cnt} that \hh binding is boosted by over 100 meV inside Ca@CNT(6,6) relative to pristine CNT(6,6) according to all the methods we consider.  

\begin{figure}[htbp]
    \centering
    \includegraphics[width=0.5\linewidth]{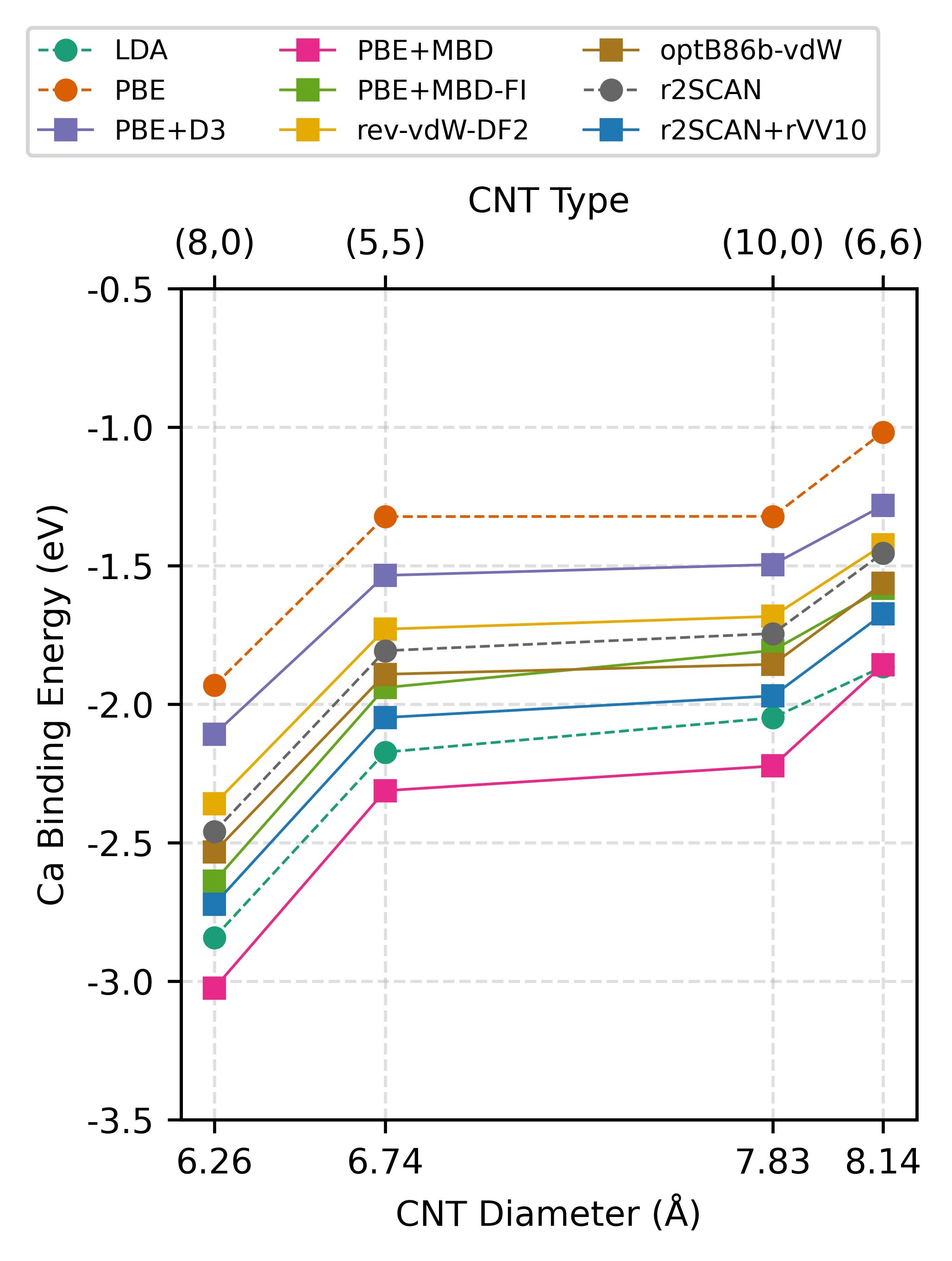}
    \caption{Ca interaction energy inside CNTs in units of eV. The chirality and diameter of the CNTs are indicated on the top and bottom x-axes, respectively. LDA, PBE and r2SCAN are shown with circles and dashed lines to distinguish them as methods not including approximations for long-range correlation.}
    \label{fig:cacnt}
\end{figure}
\begin{figure}[htbp]
    \centering
    \includegraphics[width=0.5\linewidth]{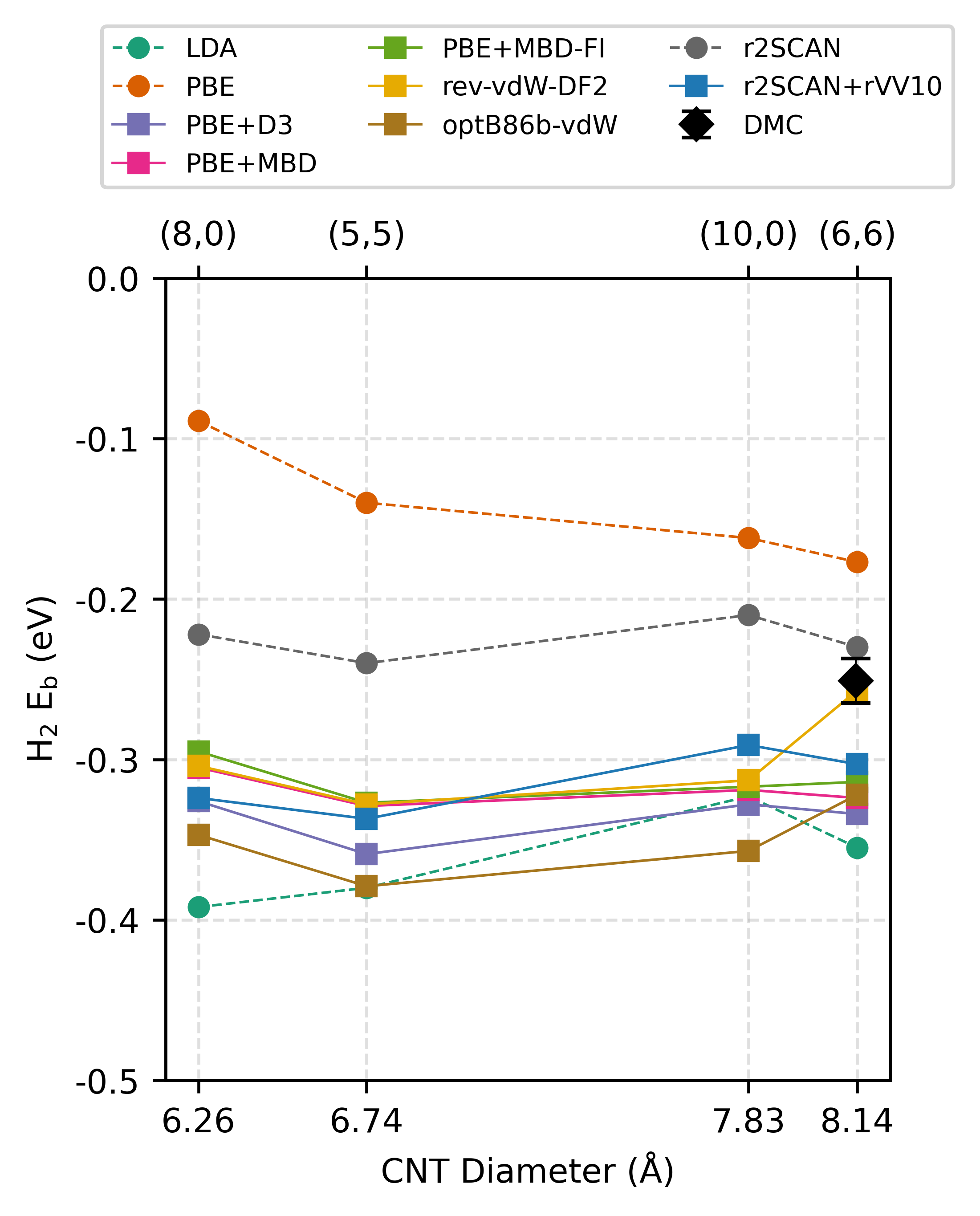}
    \caption{\hh interaction energy inside Ca@CNT systems, in units of eV. The chirality and diameter of the CNTs are indicated on the top and bottom x-axes, respectively. LDA, PBE and r2SCAN are shown with circles and dashed lines to distinguish them as methods not including approximations for long-range correlation.}
    \label{fig:h2cacnt}
\end{figure}

The Ca binding energy inside CNTs can be seen in Fig.~\ref{fig:cacnt}. The Ca binding energy is weaker for larger CNT diameters which is expected as the Ca binding energy tends towards the Ca@Gr binding energy with reduced curvature of the CNT. Fig.~\ref{fig:cacnt} also shows that the DFA predictions lie in a remarkably large range of $\sim$1 eV, indicating a large error in the absolute interaction strength. However, it can be seen that the deviations are systematic, such that the overall trend is consistent across the CNTs considered. Specifically, all DFAs we apply predict  the strength of Ca binding as CNT(8,0) > CNT(5,5) $\approx$ CNT(10,0) > CNT(6,6). Metallic CNTs, CNT(5,5) and CNT(6,6), are found to adsorb Ca more weakly than semi-conducting CNTs, CNT(8,0) and CNT(10,0), for a similar CNT diameter.  On the whole, the Ca+CNT binding energy is relatively strong and given the structural barriers to agglomeration, CNTs can be considered as an anchoring substrate for Ca adatoms.

The \hh interaction strength is predicted to be within $-0.2$ to $-0.4$ eV across all Ca@CNT systems considered here with all DFAs applied, aside from PBE. However, previous work has shown that non-local correlation has a notable impact on the \hh binding energy inside CNTs and that such configurations are susceptible to being overestimated as a result of accumulated errors in the medium-range correlation energy \cite{al-hamdani2017cnt}. Here, we report the reference DMC interaction strength for \hh+Ca@CNT(6,6) at the PBE+D3 geometry. This can be seen from Fig.~\ref{fig:h2cacnt} to be within the binding energy window of \hh storage, at $-251\pm14$ meV. As such, Ca decorated CNTs can be considered promising materials for \hh storage. 

It can be seen from Fig.~\ref{fig:h2cacnt} that rev-vdW-DF2 and r2SCAN interaction strengths agree remarkably well with DMC and yield typically weaker adsorption than the other DFAs that account for non-local correlation. Given that the geometry is fixed to be the same across all of the methods reported, we also considered the impact of optimizing each configuration for each DFA. We find that there is negligible change in the adsorption energies and structures, cementing the validity of the PBE+D3 optimized configurations and the performance of the methods (see SM~\cite{SM}). We also note that rev-vdW-DF2 and optB86b-vdW, which are both density dependent vdW-inclusive DFAs, predict a weaker adsorption on Ca@CNT(6,6) than on Ca@CNT(10,0), in contrast to the other DFAs considered. This suggests that the density-dependent vdW DFAs are more sensitive to the electronic structure of the CNT and this can be important, for example in developing data driven methods with training data from DFAs. 

\section{Discussion}\label{discussion}

The DMC references adsorption energies reported here show that DFAs typically overbind \hh and DFAs that account for long-range dispersion interactions are therefore, counterintuitively, worse. In general the selection of DFAs we consider exhibit mostly systematic shifts in the absolute adsorption energies of Ca and \hh. However, a difference in trend is found between CNT(10,0) and CNT(6,6) for rev-vdW-DF2 and optB86b-vdW against other DFAs, indicating a sensitivity to the underlying electronic structure of the zigzag and armchair nanotubes. Given the DMC reference energies reported in this work and the importance of long-range dispersion interactions in low-dimensional substrates, the rev-vdW-DF2 method is found to perform relatively well across the systems considered. 

With the DMC adsorption references reported in this work, it can be seen that DFAs that were typically applied in previous works, such as the LDA or PBE functionals, yield quantitatively unreliable but qualitatively reasonable results. 

We have reliably cemented the role of B in anchoring a Ca adatom by $\sim1.5$ eV relative to pristine graphene. In addition, Ca boosts the adsorption of \hh on the order of $\sim100$ meV relative to undecorated low-dimensional carbon materials. The \hh adsorption interaction culminates from the attractive long-range dispersion interaction between the molecule and the substrate structure as well as the local Kubas interaction with the Ca adatom. The importance of dispersion can be deduced from the net D3, MBD, MBD-FI, and rVV10 dispersion contributions in the systems considered here; these indicate 20-40$\%$ of the total interaction energy is long-range dispersion. Therefore, prospective efforts to find a suitable \hh storage material should focus on maximizing these two interactions. For example, a plethora of covalent organic frameworks have been synthesized or hypothesized which may provide a framework with higher polarizability for stronger dispersion interactions. Meanwhile, the Kubas interaction may be optimized further by considering other metal decorators such as light transition metals. The synthesis and control of single metal atom decorated low-dimensional materials remains a considerable experimental challenge \cite{Gao2024ExperimentallyStorage}, but Ca decorated graphene, for example, has been successfully synthesized using at least two different methods \cite{Emery2005Superconductivity/math,Weller2005SuperconductivityC6Ca,Chapman2016SuperconductivityLaminates}. The work presented here serves as an indication of the promising potential in these materials and support for continued efforts.

\section{Conclusion}\label{conc}
\hh molecule and Ca adatom adsorption has been reported for a selection of low-dimensional carbon allotropes (graphene and CNT based). We applied a selection of widely-used and recently developed DFAs as well as \textit{state-of-the-art} DMC to deliver robust insights. DMC and DFT methods show that boron doping in graphene (BGr) stabilizes Ca adsorption by over 1 eV compared to pristine graphene. Importantly, a small boost of 10-20 meV per \hh is found in 4H$_2$+Ca@BGr, relative to 4H$_2$+Ca@Gr, indicating that the B aided anchoring of Ca does not deteriorate the adsorption of \hh molecules and even slightly improves it. Graphene functionalized in this way makes a significant step towards the goal of obtaining adsorption energies in the range of $-200$ to $-400$ meV for viable hydrogen storage. We also computed \hh binding in Ca decorated CNTs, as it was previously shown with DMC that \hh binds inside pristine CNT(10,0) at $-115\pm14$ meV. Using Ca decoration of CNT, DMC as well as all the DFAs we consider predict \hh to bind within suitable binding energy window for hydrogen storage. Specifically, we report the DMC adsorption energy of H$_2$+Ca@CNT(6,6) at $-251\pm14$ meV. Moreover, the Ca binding energy inside CNTs with 6-8 Å diameter is stronger than $-1$ eV while the CNTs are expected to structurally inhibit Ca agglomeration.

\section*{Supplementary Material}
See the supplementary material for comprehensive details on the computational setup used in this study.

\section*{Acknowledgements}
Y.S.A., A.Z. and D.A. acknowledge support from the European Union under the Next generation EU (projects 20222FXZ33 and P2022MC742) and from Leverhulme grant no. RPG-2020-038.
Calculations were also performed using the Cambridge Service for Data Driven Discovery (CSD3) operated by the University of Cambridge Research Computing Service (www.csd3.cam.ac.uk), provided by Dell EMC and Intel using Tier-2 funding from the Engineering and Physical Sciences Research Council (capital grant EP/T022159/1 and EP/P020259/1), and DiRAC funding from the Science and Technology Facilities Council (www.dirac.ac.uk). This work also used the ARCHER UK National Supercomputing Service (https://www.archer2.ac.uk), the United Kingdom Car Parrinello (UKCP) consortium (EP/ F036884/1).


\appendix
\renewcommand{\thesection}{S\arabic{section}}
\renewcommand{\thesubsection}{S\arabic{section}.\arabic{subsection}}
\setcounter{section}{0}
\setcounter{subsection}{0}
\renewcommand{\thefigure}{S\arabic{figure}}
\renewcommand{\thetable}{S\arabic{table}}
\setcounter{figure}{0}
\setcounter{table}{0}
\clearpage
\section*{Supporting Information: Towards viable H$_2$ storage in Ca decorated low-dimensional materials with insights from reference quantum Monte Carlo}

\section{Numerical tests and details for quantum Monte Carlo}
There are a few controllable sources of error in fixed-node DMC and in this section we provide details for the time-step convergence of DMC computations and supercell tests for BGr based systems. 

\subsection{Diffusion Monte Carlo time-step convergence}
First, we report time-step convergence for $E_b^{H_2}$ on Ca@BGr and $E_b^{Ca}$ on BGr in the primitive unit cell that corresponds to a doped and decorated $(5\times5)$ graphene unit cell. Using the determinant localization approximation (DLA) we computed the FN-DMC binding energies using 0.03 and 0.01 a.u. We also performed a careful time-step analysis in our previous work on Ca@Gr~\cite{alhamdani2023jcp} that is a physically and chemically similar system. 
\begin{table}[h]
\centering
\caption{Total energies for $4\mathrm{H_2}+ \mathrm{Ca@BGr}$ and $\mathrm{Ca+BGr}$ as a function of time step.}
\label{tab:timestep_ca_bgr}
\begin{tabular}{ccccc}
\hline
Time step & $E_{4\mathrm{H_2}+ \mathrm{Ca@BGr}}$ (eV) & Error (eV) & $E_{\mathrm{Ca+BGr}}$ (eV) & Error (eV) \\
\hline
0.03 & -0.191 & 0.003 & -1.751 & 0.014 \\
0.01 & -0.188 & 0.006 & -1.771 & 0.020 \\
\hline
\end{tabular}
\end{table}

Table~\ref{tab:timestep_ca_bgr} shows agreement for the H$_2$ and Ca binding energy between the two time-steps (within the 1-$\sigma$ stochastic uncertainty). This is also consistent with our findings in previous work~\cite{alhamdani2023jcp}. As such, we use the 0.03 a.u. time-step for all further FN-DMC simulations for the BGr systems. Importantly, the values in Table~\ref{tab:timestep_ca_bgr} are not the final values reported in the main manuscript, as the supercell and spin-polarization corrections have to be accounted for also, as detailed in Sections \ref{supercell} and \ref{spincorr}.

Second, using the DLA also, we computed the FN-DMC H$_2$ binding energy in Ca@CNT(6,6) using four time steps: 0.0025, 0.005, 0.010, and 0.030 a.u. The binding energies are converged within the respective 1-$\sigma$ stochastic uncertainties for the three smallest time-steps. We can maximize the information from this data by computing the statistically weighted average, arriving at $-0.251\pm0.014$ eV as the final binding energy reported in the main results. 
\begin{table}[h]
\centering
\caption{Adsorption energy in eV of $\mathrm{H_2}$ in Ca@CNT(6,6) as a function of DMC time-step in a.u. The statistically weighted average is calculated from energies and errors from 0.0025-0.010 a.u.}
\label{tab:h2_ads_timestep}
\begin{tabular}{ccc}
\hline
Time step (a.u.)& $E_b^{H_2}$ (eV)& Error (eV) \\
\hline
0.030  & -0.291 & 0.017 \\ \hline
0.010  & -0.237 & 0.019 \\
0.005  & -0.275 & 0.023 \\
0.0025 & -0.221 & 0.063 \\
\hline
Weighted avg. & -0.251 & 0.014 \\
\hline
\end{tabular}
\end{table}

\subsection{Estimating spin polarization for QMC}\label{spincorr}
The wavefunction for QMC computations is stored in memory and therefore, depending on the hardware capability and software handling of the wavefunction in memory, it can be prohibitive to compute systems with 
large wavefunctions. In particular, the spin-polarized wavefunction is twice the size of a non-spin polarized wavefunction. While the B-spline grid density and the plane-wave cutoff used for the DFT orbitals can also be reduced to reduce the wavefunction size, doing so affects the numerical accuracy and efficiency of DMC simulations. 

We find that spin-polarization does not impact the CNT(6,6) based systems, however the BGr based systems are affected by spin-polarization due to the odd number of electrons in the unit cell with a single B dopant. We compute the net effect of spin-polarization at the DFT level for BGr systems, using different density functional approximations (LDA, PBE, and rev-vdW-DF2) and find that the spin-polarization contribution is consistent to within 5 and 6 meV agreement for $E_b^{H_2}$ and $E_b^{Ca}$, respectively. As such, we model BGr based systems using non-spin polarized orbitals and add a post-correction from DFT that is 19 meV for $E_b^{H_2}$ and $-104$ meV for $E_b^{Ca}$.

\subsection{Estimating finite size effects using twisted boundary conditions and supercell modelling}\label{supercell}
Periodic QMC simulations  carry are two important forms of finite size error (FSE). First, one-body (single-particle) FSE which arises from the discretization of allowed momentum states imposed by the finite simulation cell and periodic boundary conditions. These are analogous to Brillouin-zone sampling errors in DFT, although QMC does not use \textbf{k}-points. Twist-averaged boundary conditions reduce one-body FSE by introducing a phase in the many-body wave function across the boundaries. Averaging over twists effectively samples boundary conditions and accelerates convergence to the thermodynamic bulk limit. 

We utilized the Baldereschi point for hexagonal unit cells \cite{Baldereschi1973}, to choose an efficient set of twists for convergence. At the DFT level, a $(2\times2)$ supercell for BGr material computed at the Baldereschi point (i.e. [$\tfrac{1}{4}$,$\tfrac{1}{4}$,0]) is converged to less than $1\%$ in $E_b^{H_2}$ and $E_b^{Ca}$, as can be seen from Table~\ref{twists}. Note that we can equivalently model the Baldereschi point in a $(2\times2)$ supercell with 4 \textbf{k}-points in the primitive unit cell. 
\begin{table}[h]
\centering
\caption{Non-spin polarized LDA binding energies (eV) computed with different supercell and \textbf{k}-point choices. Percentage values denote relative differences with respect to the 9$\times$9$\times$1 \textbf{k}-mesh reference.}
\begin{tabular}{lccc}
\hline
 & 9$\times$9$\times$1 & 4 \textbf{k}-points and $(1\times1)$ & Baldereschi point and $(2\times2)$ \\
\hline
$E_{\mathrm{ads}}^{\mathrm{Ca}}$ (eV) 
    & -2.552 & -2.576 & -2.575 \\

$E_{\mathrm{ads}}^{\mathrm{H_2}}$ (eV) 
    & -0.325 & -0.325 & -0.325 \\

Relative error (\%) – Ca 
    &        & 0.93\% & 0.91\% \\

Relative error (\%) – H$_2$ 
    &        & 0.05\% & 0.05\% \\
\hline
\end{tabular}
\label{twists}
\end{table}
The excellent convergence using the Baldereschi point as applied to BGr, Ca@BGr and 4H$_2$+Ca@BGr, supports the use of the four respective twists in the primitive cell and the Baldereschi point for a single twist in the $(2\times2)$ supercells in QMC. Meanwhile, for the 1-dimensional CNT(6,6) based systems, we were able to use the same twists as the the full 9 \textbf{k}-point grid used in DFT calculations. 

Second, two-body FSE arises from artificial periodic interactions of electrons since they are modelled explicitly. Corrections such as the Kwee, Zhang, and Krakauer (KZK), Model Periodic Coulomb (MPC) and structure-factor–based methods target this contribution. The most explicit method of addressing both of the aforementioned FSEs is to simulate large unit cells, i.e. supercell calculations, to reach the thermodynamic limit. This is evidently very challenging due to the size of the system being simulated. 

We computed the KZK correction as well as supercell calculations for the BGr based systems and the results can be found in Table~\ref{fse}.
\begin{table}[h]
\centering
\caption{Finite-size error (FSE) corrections using the KZK scheme for different systems and supercell choices. Energies are in eV.}
\begin{tabular}{lcccc}
\hline
 & 4H$_2$+Ca@Gr & 4H$_2$+Ca@BGr & Ca+Gr & Ca+BGr \\
\hline
KZK-(1$\times$1) 
    & 0.011  & 0.019  & 0.124  & 0.101 \\

KZK-(2$\times$2)
    &        & 0.005  &        & 0.025 \\
\hline
\end{tabular}
\label{fse}
\end{table}

We use the data that incurs the smallest FSE to compute the best possible binding energies. As the FSE is smaller in the $(2\times2)$ BGr supercells, we correct $E_b^{H_2}$ by 5 meV and $E_b^{Ca}$ by 25 meV. In the case of Gr, we only have the correction for the  $(1\times1)$ supercell, and therefore we use 11 meV and 124 meV to correct $E_b^{H_2}$ by 5 meV and $E_b^{Ca}$, respectively.

\section{Computing the energy barrier for Ca diffusion across BGr}
\begin{figure}[htbp]
    \centering
    \includegraphics[width=0.5\linewidth]{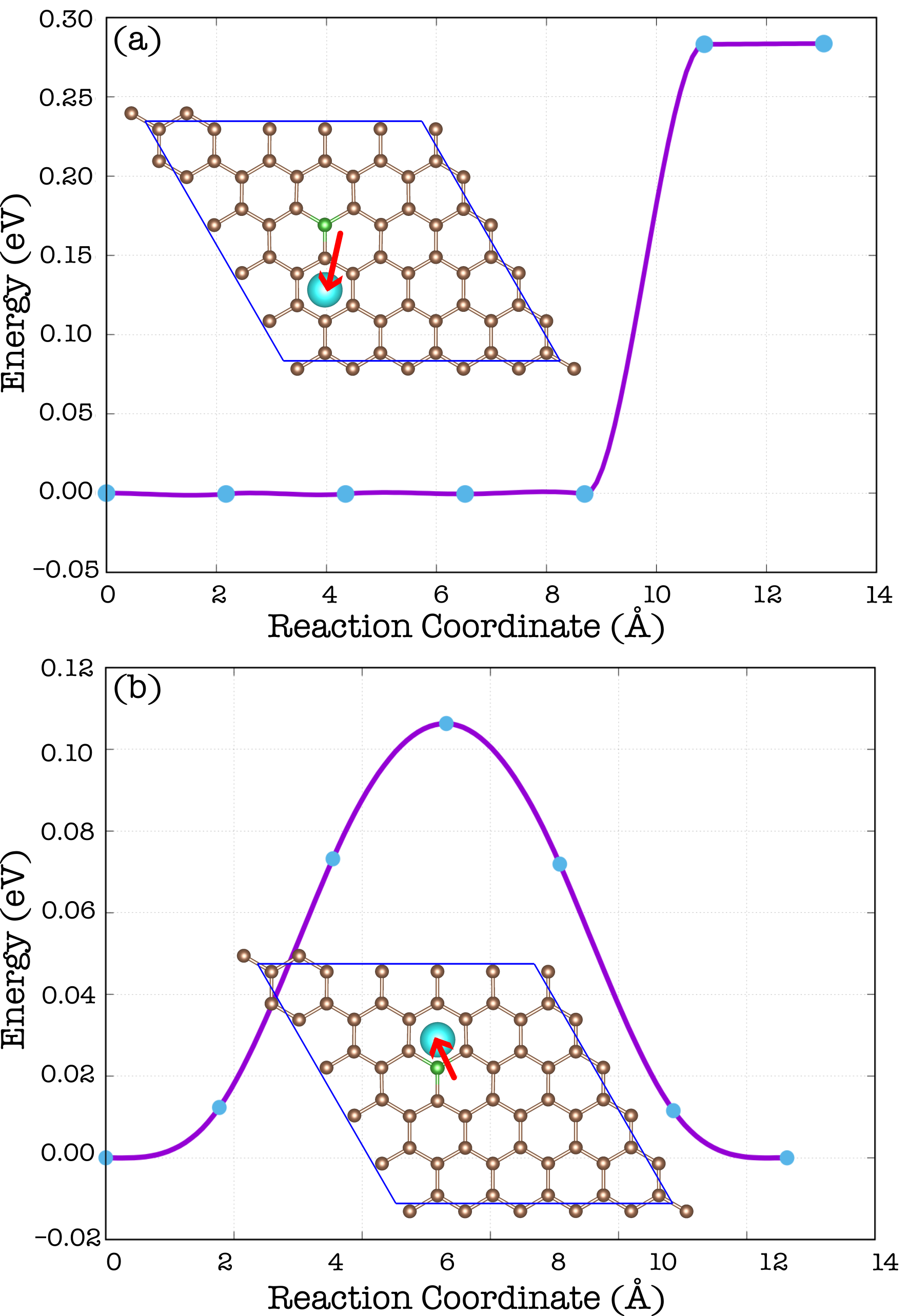}
    \caption{The NEB pathway for two pathways: (a) Ca diffusion from BC$_5$ ring to nearest C$_6$ ring, and (b) the Ca diffusion from a BC$_5$ ring to the next nearest BC$_5$ ring. The pathways are shown by red arrows in the insets.}\label{neb}
\end{figure}

The Ca diffusion barrier was computed using the climbing image nudged elastic band (NEB) method with 5 replicas and a spring force constant of 5 eV \AA$^{-2}$ with nudging ~\cite{Henkelman, Henkelman2000climb, Henkelman2000imp}. Two paths have been considered as shown in Fig.~\ref{neb}. 

\section{Optimized H$_2$+Ca@CNT systems with different density functional approximations}
\begin{figure}[htbp]
    \centering
    \includegraphics[width=0.5\linewidth]{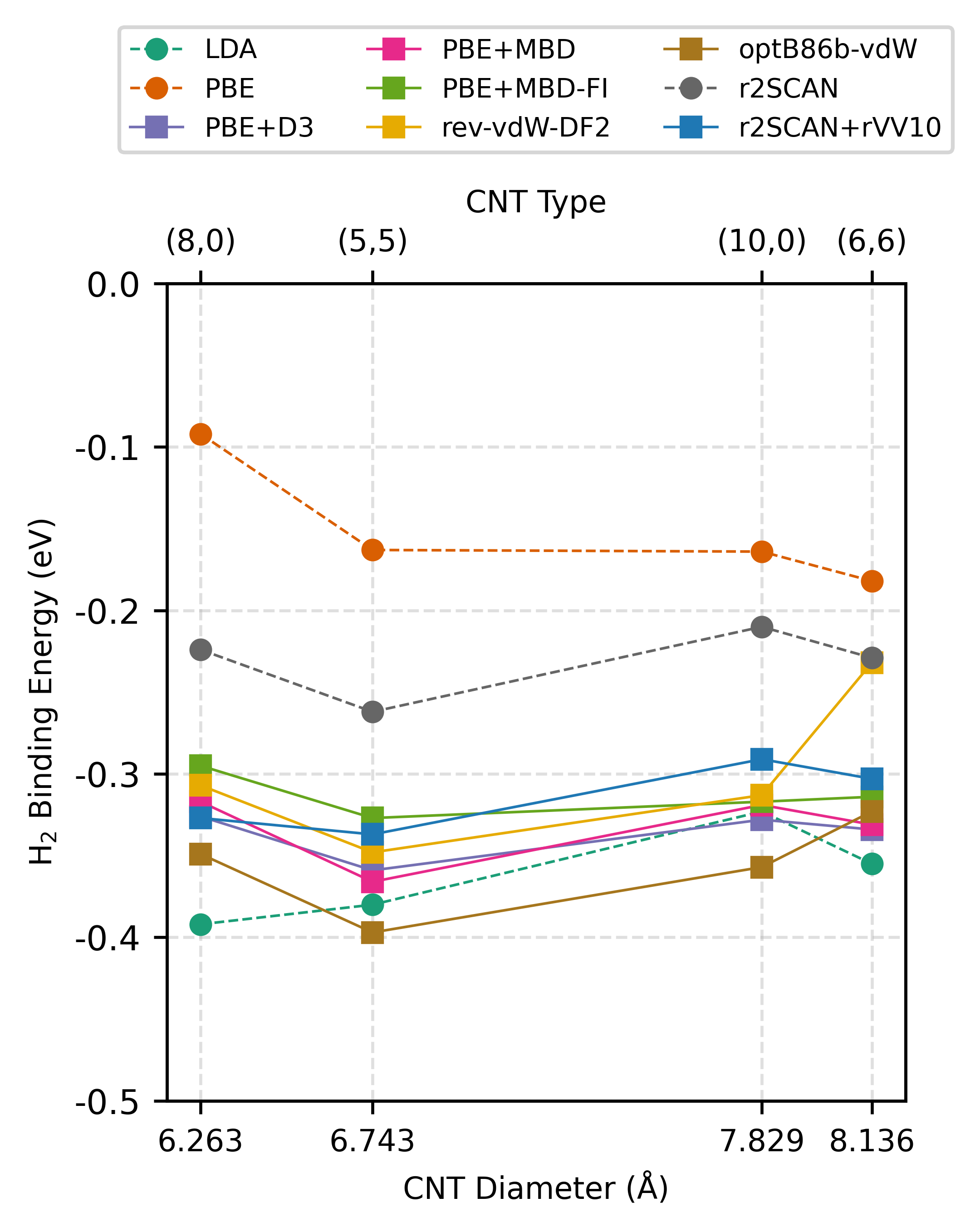}
    \caption{H$_2$ binding energy on Ca@CNT materials with different density functional approximations after full geometry relaxation of the configurations.}\label{h2cacntopt}
\end{figure}

\begin{figure}[htbp]
    \centering
    \includegraphics[width=0.5\linewidth]{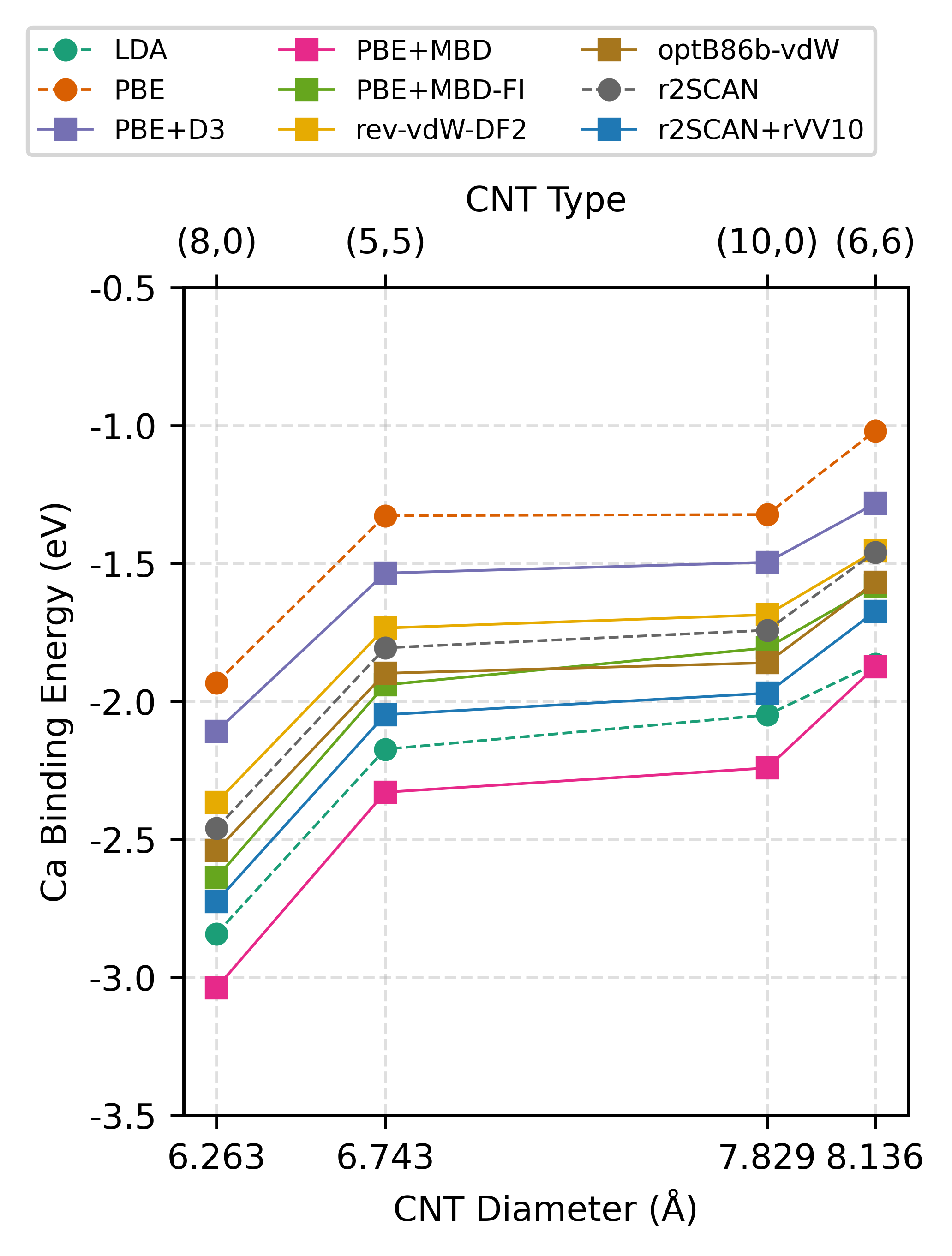}
    \caption{Ca binding energy inside CNT materials with different density functional approximations after full geometry relaxation of the configurations.}\label{cacntopt}
\end{figure}

We report the fully geometry optimized binding energies for H$_2$ on Ca@CNT materials and Ca inside CNTs with a selection of density functional approximations in Fig.~\ref{h2cacntopt} and Fig.~\ref{cacntopt}, respectively. We applied consistent numerical thresholds across all calculations, as is applied with PBE+D3. The adsorption strengths are mildly affected and the overall trends remain consistent.


\bibliography{references,methods}
\bibliographystyle{ieeetr}

\end{document}